\documentclass[prd,aps,nofootinbib,showpacs, 11 pt]{revtex4}
\usepackage{amsmath}
\usepackage{amsmath,graphicx,color,epsfig}

\begin{document}
\title{Exclusive $B_s$ decays to the charmed mesons $D_s^+(1968,2317)$ in the standard model}
\author{Run-Hui Li$^{a,b}$, Cai-Dian L\"u$^{a}$ and Yu-Ming Wang$^{a}$}

\affiliation{
 \it $^a$ Institute of High Energy Physics and Theoretical Physics Center for Science Facilities,
  P.O. Box 918(4) Beijing 100049,  China\\
 \it $^b$  School of Physics, Shandong University, Jinan 250100, China }

\begin{abstract}

The transition form factors of  $\bar B_s^0\to D_{s}^+(2317)$ and
$\bar B_s^0\to D_s^+(1968)$ at large recoil region are investigated
in the light cone sum rules approach, where the heavy quark
effective theory is adopted to describe the form factors at small
recoil region. With the form factors obtained, we carry out a
detailed analysis on both the semileptonic decays $\bar B_s^0\to
D_s^+(1968,2317) l \bar{\nu}_l$ and nonleptonic decays  $B_s \to
D_s^+(1968,2317)  M$ with $M$ being a light meson or a charmed meson
under the factorization approach. Our results show that the
branching fraction of $\bar B_s^0\to D_s^+(2317) \mu \bar{\nu}_\mu$
is around $2.3 \times 10^{-3}$, which should be detectable with ease
at the Tevatron and LHC. It is also found that the branching
fractions of $\bar B_s^0\to D_s^+(1968) l \bar{\nu}_l$ are almost
one order larger than those of the corresponding $B_s^0\to
D_s^+(2317) l \bar{\nu}_l$ decays.  The consistency of predictions
for $B_s \to D_s^+(1968,2317) L$ ($L$ denotes a light meson) in the
factorization assumption and $k_T$ factorization also supports the
success of color transparency mechanism in the color allowed decay
modes. Most two-charmed meson decays of $B_s$ meson possess quite
large branching ratios that  are  accessible in the experiments.
These channels are of great importance to explore the hadronic
structure of charmed mesons as well as the nonperturbative dynamics
of QCD.

\end{abstract}
\pacs {14.40.Lb, 13.20.He, 11.55.Hx}

\maketitle
\section{Introduction}

Enthusiasm for the open charm spectroscopy has been renewed since
the announcement of a narrow low mass state $D_s(2317)$ with
unexpected and intriguing prosperities, observed in the $D_s \pi^0$
decay mode by BaBar collaboration\cite{Aubert:2003fg}. The analysis
of these charmed resonances can be considerably simplified in the
limit of infinite heavy quark mass, when the heavy quark acts as a
static color source so that its spin is decoupled from the total
angular momentum of the residual light degrees of freedom. Weak
production of charmed mesons in the $B_s$ meson decays induced by
the $b \to c$ transition serves as an ideal platform to scrutinize
the KM mechanism of the standard model (SM), explore the dynamics of
strong interactions as well as probe the signals of new physics.
Moreover, valuable information on the inner structures of the exotic
charmed mesons can also be extracted from the rare decays realized
via the $b \to c$ transition.

On the experimental aspect, $B_s$ meson will be copiously
accumulated at the LHC, which makes the investigations of the
$B_s$'s static prosperities and its decay characters promising. On
the theoretical side, the heavy quark symmetry can put stringent
constraint on the form factors responsible for $\bar B_s^0\to
D_s^+(1968,2317)$ transition. As for the $B_s$ meson transitions to
the lowest lying charmed mesons, one needs to introduce a universal
Isgur-Wise function $\xi(v \cdot v^{\prime})$, whose normalization
is $\xi(v \cdot v^{\prime}=1)=1$ as a consequence of  the flavor
conserving vector current. However, the heavy quark symmetry could
not predict the normalization of  the universal form factor
$\tau_{1/2}$ responsible for the decays of $B_q$ meson to the
doublet $J_{s_l}^P=(0^+, 1^+)_{1/2}$ \cite{DeFazio:2000up},
therefore one has to rely on some nonperturbative methods to deal
with the $\bar B_s^0 \to D_s^+(1968,2317)$ transition form factors.

Currently, there have been some studies on the semileptonic decays
$\bar B_s^0\to D_s^+(1968,2317) l \bar{\nu}_l$ ranging from
phenomenological model \cite{Zhao:2006at} to QCD sum rules approach
\cite{Huang:2004et,Aliev:2006qy,Azizi:2008tt},  PQCD approach
\cite{Li:2008ts} and Lattice QCD \cite{Hashimoto:1999yp,de
Divitiis:2007ui,de Divitiis:2007uk}. It could be found that the
available theoretical predictions vary from each other, hence the
investigation of these modes in the framework that is  well  rooted
in the quantum field theory is in demand.

Light cone sum rule(LCSR) offers an systematic way to compute the
soft contribution to the transition form factor almost
model-independently\cite{LCSR 1, LCSR 2,LCSR 3,LCSR 4,LCSR 5}. As a
marriage of the standard QCD  sum rule (QCDSR) technique \cite{SVZ
1, SVZ 2,SVZ 3} and the theory of hard exclusive process, LCSR cures
the problem of QCDSR applying to the large momentum transfer by
performing the operator product expansion (OPE) in terms of the
twists of revelent operators rather than their
dimensions~\cite{braun talk}. Therefore, the principal discrepancy
between QCDSR and LCSR consists in that non-perturbative vacuum
condensates representing the long-distance quark and gluon
interactions in the short-distance expansion are substituted by the
light cone distribution amplitudes (LCDAs) describing the
distribution of longitudinal momentum carried by the valence quarks
of hadronic bound system in the expansion of transverse-distance
between partons in the infinite momentum frame. Phenomenologically,
LCSR has been applied widely to the investigation of the transition
of mesons and baryons in recent years \cite{LCSR a1, LCSR a2, LCSR
a3, LCSR a4, LCSR a5, LCSR a6, LCSR a7, LCSR a8, LCSR a9}.

In this work, we will employ the LCSR approach to compute the $\bar
B_s^0 \to D_s^+(1968,2317)$ form factors, and then analyze the
mentioned semileptonic modes as well as the nonleptonic decays $B_s
\to D_{sJ} M$, with $M$ being a light meson or a charmed meson,
under the factorization approach. It is expected that we can win the
double benefit from such decays:  gain better understanding on the
dynamics of strong interactions  and clarify the inner structures of
$D_s^+(1968,2317)$ mesons.

The layout of this paper is as follows: We firstly collect the
distribution amplitudes of $D_s^+(1968,2317)$ mesons in the section
II. The equation of motion and heavy quark symmetry are employed to
simplify the structures of hadronic wavefunctions. The sum rules for
the transition form factors  $\bar B_s^0 \to D_s^+(1968,2317)$ up to
twist-3 are then derived in section III, where the relation of form
factors in the heavy quark limit are found to be well respected in
the LCSR approach. The numerical analysis of LCSR for the transition
form factors  at large recoil region are displayed in section IV.
Heavy quark effective theory (HQET) is adopted to describe the $\bar
B_s^0 \to D_s^+(1968,2317)$ transitions at the small recoil region.
Moreover, detailed comparisons between  the form factors obtained
under various approaches are also presented here.  Utilizing these
form factors,  the  branching fractions of semileptonic decays $\bar
B_s^0\to D_s^+(1968,2317) l \bar{\nu}_l$ and nonleptonic decays $B_s
\to D_s^+(1968,2317)  M$ are calculated in section V. In particular,
some remarks on the factorization of nonleptonic modes are given
here. The last section is devoted to the conclusion.

\section{Effective Hamiltonian  and Light cone distribution amplitudes}

\subsection{Effective Hamiltonian for the $b$ quark decays }

In this subsection, we would like to collect the effective
Hamiltonian for $b$ quark decays after integrating out the particles
including top quark, $W^{\pm}$ and $Z$ bosons above scale
$\mu=O(m_b)$.  For the semileptonic $b\to cl\bar \nu_l$ transition,
the effective Hamiltonian can be written as
 \begin{eqnarray}
 {\cal H}_{eff}(b\to cl\bar \nu_l)=\frac{G_F}{\sqrt{2}}V_{cb}\bar
 c\gamma_{\mu}(1-\gamma_5)b \bar l\gamma^{\mu}(1-\gamma_5)\nu_l.
 \end{eqnarray}
For the nonleptonic transition with $\Delta B=1$, the effective
Hamiltonian is specified as
\begin{eqnarray}
{\cal H}_{\rm eff} = \frac{G_F}{\sqrt2} \sum_{p=u,c} \!
   \lambda_p^{(D)} \bigg( C_1\,Q_1^p + C_2\,Q_2^p
   + \!\sum_{i=3,\dots, 10}\! C_i\,Q_i + C_{7\gamma}\,Q_{7\gamma}
   + C_{8g}\,Q_{8g} \bigg) + \mbox{h.c.} \,, \nonumber \\
\label{HeffSM}
\end{eqnarray}
where the CKM factors are
\begin{eqnarray}
\lambda_p^{(D)}\equiv V_{pb}V^*_{pD}= \left\{
\begin{array}{l}
 V_{pb}V_{pd}^{\ast}, \qquad \mbox{ { for $b \to d$ transition} ; }
\\
V_{pb}V_{ps}^{\ast},  \qquad \mbox{ { for $b \to s$ transition} .}
\end{array}
\right.
\end{eqnarray}
The function $Q_i$ are the local four-quark operators:
\begin{itemize}
\item
current-current (tree) operators£º
\begin{eqnarray}
 Q_1^p = (\bar p b)_{V-A} (\bar D p)_{V-A} \,, \qquad
    Q^p_2 = (\bar p_i b_j)_{V-A} (\bar D_j p_i)_{V-A} \,,
\end{eqnarray}
\item
QCD penguin operators:
\begin{eqnarray}
 Q_3 &=& (\bar D b)_{V-A} \sum{}_{\!q}\,(\bar q q)_{V-A}
\,,
    \hspace{1.7cm}
    Q_4 = (\bar D_i b_j)_{V-A} \sum{}_{\!q}\,(\bar q_j q_i)_{V-A} \,,
    \nonumber\\
   Q_5 &=& (\bar D b)_{V-A} \sum{}_{\!q}\,(\bar q q)_{V+A} \,,
    \hspace{1.7cm}
    Q_6 = (\bar D_i b_j)_{V-A} \sum{}_{\!q}\,(\bar q_j q_i)_{V+A}
    \,,
\end{eqnarray}
\item
electro-weak penguin operators:
\begin{eqnarray}
 Q_7 &=& (\bar D b)_{V-A}
\sum{}_{\!q}\,{\textstyle\frac32} e_q
    (\bar q q)_{V+A} \,, \hspace{1.11cm}
    Q_8 = (\bar D_i b_j)_{V-A} \sum{}_{\!q}\,{\textstyle\frac32} e_q
    (\bar q_j q_i)_{V+A} \,, \nonumber \\
   Q_9 &=& (\bar D b)_{V-A} \sum{}_{\!q}\,{\textstyle\frac32} e_q
    (\bar q q)_{V-A} \,, \hspace{0.98cm}
    Q_{10} = (\bar D_i b_j)_{V-A} \sum{}_{\!q}\,{\textstyle\frac32} e_q
    (\bar q_j q_i)_{V-A} \,, \,\,\,\,\,
 \end{eqnarray}
\item electromagnetic and chromomagnetic dipole operators  :
 \begin{eqnarray}
 Q_{7\gamma} &=& \frac{-e}{8\pi^2}\,m_b\,
    \bar D\sigma_{\mu\nu}(1+\gamma_5) F^{\mu\nu} b \,,
    \hspace{0.81cm}
   Q_{8g} = \frac{-g_s}{8\pi^2}\,m_b\,
    \bar D\sigma_{\mu\nu}(1+\gamma_5) G^{\mu\nu} b \,,
     \end{eqnarray}
\end{itemize}
where $i$ and $j$ are the color indices,  $(\bar{q}_1 q_2)_{V\pm
A}=\bar{q}_1 \gamma_\mu(1\pm\gamma_5)q_2$ and the sum runs  over all
active quark flavors in the effective theory, i.e., $q=u,d,s,c,b$.
The combinations $a_i$ of Wilson coefficients are defined as usual
\cite{Ali:1998eb}:
\begin{eqnarray}
a_1= C_2+C_1/3, & a_3= C_3+C_4/3,~a_5= C_5+C_6/3,~a_7=
C_7+C_8/3,~a_9= C_9+C_{10}/3,\nonumber \\
 a_2= C_1+C_2/3, & a_4=
C_4+C_3/3,~a_6= C_6+C_5/3,~a_8= C_8+C_7/3,~a_{10}= C_{10}+C_{9}/3.
\end{eqnarray}

\subsection{Distribution amplitudes of $D_s^+(1968)$}

The distribution amplitudes of pseudoscalar meson $D_s^+$ can be
defined as \cite{Kurimoto:2002sb}
\begin{eqnarray}
 \langle D_s^+(P)|\bar c(y)\gamma_{\mu}\gamma_5 s(w)|0\rangle
 &=&-if_{D_s}p_{\mu} \int_0^1 du e^{-i(P-k)\cdot y-ik\cdot w}\phi_D^v(u)
 \nonumber  \\
&&  -\frac{i}{2}f_{D_s}m_{D_s}^2\frac{z_{\mu}}{P\cdot z}
 \int_0^1du e^{-i(P-k)\cdot y-ik\cdot w}g_D(u), \nonumber\\
 \langle D_s^+(P)|\bar c(y)\gamma_5 s(w)|0\rangle
 &=&-i m_0 f_{D_s} \int_0^1 du e^{-i(P-k)\cdot y-ik\cdot
 w}\phi_D^p(u),\nonumber\\
 \langle D_s^+(P)|\bar c(y)\sigma_{\mu\nu}\gamma_5 s(w)|0\rangle
 &=&\frac{i}{6}f_{D_s}m_0(1-\frac{m_{D_s}^2}{m_0^2})(P_{\mu}z_{\nu}-P_{\nu}z_{\mu})
 \int_0^1 du e^{-i(P-k)\cdot y-ik\cdot
 w}\phi_D^{\sigma}(u),\label{eq:Dnonmatrix}
\end{eqnarray}
where $z=y-w$ and  $u=1- \frac{k^+}{P^+}$ is the longitudinal
momentum fraction carried by the charm quark. In the heavy quark
limit, the chiral  mass can be simplified as
\begin{eqnarray}
m_0=\frac{m^2_{D_s}}{m_c+m_s}=m_{D_s}+O(\bar\Lambda),
\end{eqnarray}
which indicates that the  contribution from the distribution
amplitude $\phi_D^{\sigma}(u)$ is suppressed by
$O(\bar\Lambda/m_{D_s})$ compared with that from $\phi_D^{v}(u)$ and
$\phi_D^{p}(u)$. It can also be observed that the twist-4
distribution amplitude $g_D(u)$ contributes at the power of $r^2$
with $r=\frac{m_{D_s}}{m_{B_s}}$, therefore it can be safely
neglected in the numerical calculations.

In the next place, we would like to derive the relations between the
distribution amplitudes $\phi_D^{v}(u)$ and $\phi_D^{p}(u)$ in the
heavy quark limit with the help of the equation of motion. Following
the Ref.~\cite{Kurimoto:2002sb}, the nonlocal matrix element with
the insertion of  pseudotensor current can be rewritten as
\begin{eqnarray}
 \langle D_s^+(P)|\bar c(y)\sigma_{\mu\nu}\gamma_5 s(w)|0\rangle
 =\langle D_s^+(P)|\bar c(y)\gamma_{\mu}\gamma_{\nu}\gamma_5 s(w)|0\rangle
 -ig_{\mu\nu}\langle D_s^+(P)|\bar c(y)\gamma_5 s(w)|0\rangle.
\end{eqnarray}
Differentiating both sides of the above equation with respect to
$w_{\nu}$ for $\mu=-$ and to $y_{\mu}$ for $\nu=+$, we have
\begin{eqnarray}
\label{pesudoscalar wf 1}
\int du \bar u\phi_D^p(u)e^{-i(P-k)\cdot
y-ik\cdot
w}=O(\bar\Lambda/m_{D_s}),\label{eq:phiDp}\\
\int du [\phi_D^p(u)-\phi_D^v(u)]e^{-i(P-k)\cdot y-ik\cdot
w}=O(\bar\Lambda/m_{D_s}), \label{pesudoscalar wf 2}
\end{eqnarray}
with $\bar{u}\equiv 1-u$. As shown in Eq. (\ref{pesudoscalar wf 1}),
the distribution amplitude $\phi_D^p$ peaks at the region of $\bar u
\sim O(\bar{\Lambda}/m_{D_s})$. Eq. (\ref{pesudoscalar wf 2})
indicates that the distribution amplitudes $\phi_D^{v}(u)$ and
$\phi_D^{p}(u)$  have the same normalizations
\begin{eqnarray}
\int_0^1 du \phi_D^{v}(u)=\int_0^1 du \phi_D^{p}(u) \equiv \int_0^1
du \phi_D(u)=f_{D_s}.
\end{eqnarray}
In this way, one can express the nonlocal matrix elements relevant
to the pseudoscalar $D_s$ meson in the heavy quark limit as
\begin{eqnarray}
\langle D^+_s(P)|\bar c(z)_l s(0)_j|0\rangle= \frac{i} { 4}\int_0^1
du e^{iuP \cdot z}\phi_D(u)[\gamma_5(\not{P}+m_{D_s})]_{jl},
\end{eqnarray}
The model of  $\phi_D(u)$ adopted in this work is
\begin{eqnarray}
 \phi_D(u)=f_{D_s}6u (1-u)[1-C_D(1-2u)],
\end{eqnarray}
where the shape parameter $C_D=0.78$ is determined to fit the
requirement that $\phi_D(u)$ has a maximum at $\bar
u=\frac{m_{D_s}-m_c}{m_{D_s}}$.

\subsection{Distribution amplitudes of $D_s^+(2317)$}

Following the same philosophy, the distribution amplitudes of scalar
charmed meson $D^{\ast}_{s0}$  can be defined by \cite{Chen:2003rt}
\begin{eqnarray}
 \langle D_{s0}^+(2317)(P)| \bar c(z)_j s(0)_l |0\rangle=\frac{1}{4}
 \int_0^1 du e^{iuP\cdot z}\{-(\not{P})_{lj}\Phi_{D1}(u) +
 m_{D_{s0}}(I)_{lj}\Phi_{D2}(u)\},
 \label{DAs of Ds(2317)}
\end{eqnarray}
where $D^{\ast}_{s0}$ denotes  the $D_s^+(2317)$ meson and the
normalizations of distribution amplitudes are
 \begin{eqnarray}
 \int_0^1 du\Phi_{D1}(u)=f_{D_{s0}^{\ast}}\;,\;
 \int_0^1 du\Phi_{D2}(u)=\widetilde{f}_{D_{s0}^{\ast}}.
 \end{eqnarray}
 The decay constants $f_{D_{s0}^{\ast}}$ and $\widetilde{f}_{D_{s0}^{\ast}}$
 are given by
 \begin{eqnarray}
 \langle 0|\bar s \gamma_{\mu}
 c|D^{\ast}_{s0}(P)\rangle=f_{D_{s0}^{\ast}}P_{\mu}\;,\;
 \langle 0|\bar s
 c|D_{s0}(P)\rangle=m_{D_{s0}}\widetilde{f}_{D_{s0}^{\ast}},
\end{eqnarray}
where  $f_{D^*_{s0}}=(m_c-m_s)\widetilde{f}_{D^*_{s0}}/m_{D^*_{s0}}$
with $m_c$ and $m_s$ being the current masses of charm quark and
strange quark, respectively.

Again, with the help of equation of motion,  one can find that the
distribution amplitudes $\Phi_{D1}(u)$ and $\Phi_{D2}(u)$ differ at
the order of $\bar{\Lambda}/m_{D_{s0}^{\ast}} \sim
(m_{D_{s0}^{\ast}}-m_c) / m_{D_{s0}^{\ast}}$. Hence, for the leading
power calculation, it is reasonable to parameterize the distribution
amplitudes $\Phi_{D1}(u)$ and $\Phi_{D2}(u)$ in the following form
\begin{eqnarray}
 \Phi_{D1}(u)=\Phi_{D2}(u)=\widetilde{f}_{D^*_{s0}}6u (1-u) [1 + a(1-2u)]\;
\end{eqnarray}
in the heavy quark limit. $\widetilde{f}_{D^*_{s0}}=(225\pm 25)
\rm{MeV}$ has been determined from the two-point QCD sum rules. The
shape parameter $a=-0.21$ is fixed under the condition that the
distribution amplitudes $\Phi_{D_i}(u)$ possess the maximum  at
$\bar u=\frac{m_{D_{s0}^{\ast}}-m_c}{m_{D_{s0}^{\ast}}}$ with the
charm quark mass $m_c=1.275 \rm{GeV}$. It is worthwhile to point out
that the intrinsic $b$ dependence of the charmed meson distribution
amplitudes has been neglected in the above analysis, which will
introduce more free parameters.

\section{Light cone sum rules for form factors}

\subsection{Sum rules for $\bar B_s^0\to D_{s}^+(2317)$ transition form factors}

The hadronic matrix element involved in the $\bar B_s^0\to
D_{s0}^{\ast +}$ transition can be  parameterized as
\begin{eqnarray}
 \langle D_{s0}^{\ast +}(P)|\bar c\gamma_{\mu}\gamma_5 b|\bar B_s(P+q)\rangle
 =-i[f_{D^*_{s0}}^+(q^2)P_{\mu}+f_{D^*_{s0}}^-(q^2)q_{\mu}].\label{eq:form1}
\end{eqnarray}
Following the standard procedure of  sum rules, the correlation
function for $f_{D_{s0}^{\ast}}^+(q^2)$ and
$f_{D_{s0}^{\ast}}^-(q^2)$ is chosen as
\begin{eqnarray}
 \Pi_{\mu}(P,q)=-\int d^4x e^{iq\cdot x}\langle
 D^{\ast +}_{s0}(P)|T\{j_{2\mu}(x),j_1(0)\}|0\rangle,\label{eq:cofun1}
\end{eqnarray}
where the current $j_{2\mu}(x)=\bar c(x)\gamma_{\mu}\gamma_5 b(x)$
describes the $b\to c$ weak transition and $j_1(0)=\bar
b(0)i\gamma_5 s(0)$ denotes the $\bar B_s$ channel.

Inserting the complete set of states between the currents in Eq.
(\ref{eq:cofun1}) with the same quantum numbers as $B_s$, we can
arrive at the hadronic representation of the correlation function
\begin{eqnarray}
\Pi_{\mu}(P,q)&=&i\frac{\langle D_{s0}^{*+}(P)|\bar
c(0)\gamma_{\mu}\gamma_5b(0)|\bar B_s(P+q)\rangle \langle
B_s(P+q)|\bar b(0)i\gamma_5
s(0)|0\rangle}{m^2_{B_s}-(P+q)^2}\nonumber\\
&&+\sum_h i\frac{\langle D_{s0}^{*+}(P)|\bar
c(0)\gamma_{\mu}\gamma_5b(0)|\bar h(P+q)\rangle \langle h(P+q)|\bar
b(0)i\gamma_5 s(0)|0\rangle}{m^2_{h}-(P+q)^2},\label{eq:cofunh}
\end{eqnarray}
where the definition of $B_s$ meson decay constant is
 \begin{eqnarray}
 \langle B_s|\bar b i \gamma_5
 s|0\rangle=\frac{m^2_{B_s}}{m_b+m_s}f_{B_s}.\label{eq:BsM}
\end{eqnarray}
Combining (\ref{eq:form1}), (\ref{eq:BsM}) and (\ref{eq:cofunh}), we
have
\begin{eqnarray}
 \Pi_{\mu}(P,q)&=&\frac{m^2_{B_s}f_{B_s}}{(m_b+m_s)[m^2_{B_s}-(P+q)^2]}[f_{D^*_{s0}}^+(q^2)P_{\mu}+f_{D^*_{s0}}^-(q^2)q_{\mu}]
 \nonumber \\
 &&+\int_{s_0^{\bar B_s}}^{\infty}ds
 \frac{\rho_+^h(s,q^2)P_{\mu}+\rho_-^h(s,q^2)q_{\mu}}{s-(P+q)^2},
\end{eqnarray}
where we have expressed the contributions from higher states of the
$B_{s}$ channel in the form of dispersion integral with
$s_0^{B_{s}}$ being the threshold parameter corresponding to the
$B_{s}$ channel.

On the theoretical side, the correlation function (\ref{eq:cofun1})
can  be also calculated in the perturbative theory with the help of
the OPE  technique at the deep Euclidean region $P^2,q^2=-Q^2\ll 0$:
\begin{eqnarray}
 \Pi_{\mu}(P,q)&=&\Pi_+^{\rm{QCD}}(q^2,(P+q)^2)P_{\mu}+\Pi_-^{\rm{QCD}}(q^2,(P+q)^2)q_{\mu} \\
 &=&\int_{(m_b+m_s)^2}^{\infty}ds\frac{1}{\pi}\frac{\rm{Im}\Pi_+^{\rm{QCD}}(q^2,(P+q)^2)}{s-(P+q)^2}P_{\mu}
 +\int_{(m_b+m_s)^2}^{\infty}ds\frac{1}{\pi}\frac{\rm{Im}\Pi_-^{\rm{QCD}}(q^2,(P+q)^2)}{s-(P+q)^2}q_{\mu}. \nonumber \label{eq:Piquark}
\end{eqnarray}
Making use of the quark-hadron duality
\begin{eqnarray}
 \rho_i^h(s,q^2)=\frac{1}{\pi}{\rm{Im}}\Pi_i^{\rm{QCD}}(q^2,(P+q)^2)\Theta(s-s_0^h),\label{eq:rhospec}
\end{eqnarray}
with $i=``+,-"$ and  performing Borel transformation on both sides
of Eq. (\ref{eq:rhospec}) with respect to $(P+q)^2$,  the sum rules
for the form factors can be written as
\begin{eqnarray}
 f_i(q^2)=\frac{m_b+m_s}{\pi
 f_{B_s}m^2_{B_s}}\int_{(m_b+m_s)^2}^{\infty}ds\;{\rm{Im}}\Pi_i^{\rm{QCD}}(q^2,s){\rm exp}(\frac{m^2_{B_s}-s}{M^2}).
 \label{sum rules for Ds(2317)}
\end{eqnarray}

To the leading order of $\alpha_s$, the correlation function can be
calculated by contracting the bottom quark fields in Eq.
(\ref{eq:cofun1}) and inserting the free $b$ quark propagator
\begin{eqnarray}
 \Pi_{\mu}(P,q)=i \int d^4x\int\frac{d^4k}{(2\pi)^4}\frac{e^{i(q-k)\cdot x}}{m_b^2-k^2}
 \langle D_{s0}^{\ast +}(P)|\bar
 c(x)\gamma_{\mu}\gamma_5(\not{k}+m_b)i\gamma_5
 s(0)|0\rangle.\label{eq:Piquark2}
\end{eqnarray}
It should be pointed out that the full quark propagator also
receives corrections from the background field
\cite{cite:background1,cite:background2}, which can be written as
\begin{eqnarray}
\langle 0| T \{{b_i(x) \bar{b}_j(0)}\}| 0\rangle &=& \delta_{ij}\int
{d^4 k \over (2 \pi)^4} e^{-i kx}{i \over \not \! k -m_b} -i g \int
{d^4 k \over (2 \pi)^4} e^{-i kx} \int_0^1 dv  [{1 \over 2} {\not k
+m_b \over (m_b^2 -k^2)^2} G^{\mu \nu}_{ij}(v x)\sigma_{\mu \nu }\nonumber \\
&& +{1 \over m_b^2-k^2}v x_{\mu} G^{\mu \nu}(v x)\gamma_{\nu}],
\end{eqnarray}
where the first term is the free-quark propagator and $G^{\mu
\nu}_{i j}=G_{\mu \nu}^{a} T^a_{ij}$ with ${\mbox{Tr}}[T^a T^b]={1
\over 2}\delta^{ab}$. Substituting the second term  proportional to
the gluon field strength into the correlation function can result in
the distribution amplitudes corresponding to the higher Fock states
of $ D_{s}^+(2317)$ meson. It is expected that such corrections
associating with the LCDAs of higher Fock states do not play any
significant roles in the sum rules for transition form factors
\cite{higher Fock state}, and  hence can be safely neglected.

Substituting Eq. (\ref{DAs of Ds(2317)}) into Eq.
(\ref{eq:Piquark2}) and performing the integral in the coordinate
space, the correlation function in the momentum representation at
the quark level can be written as
\begin{eqnarray}
\Pi _{\mu }\left(P,q\right) = \int_{u_0}^1 \frac{du}{u}
\big[(m_b+um_{D_{s0}})\phi_{D^*_{s0}}(u)P_{\mu}+m_{D_{s0}}\phi_{D^*_{s0}}(u)q_{\mu}\big]
e^{-(m_b^2 - \bar u q^2 + u\bar u P^2)/(u M^2)}, \label{QCD
representation of correlation 1}
\end{eqnarray}
with
\begin{eqnarray}
 u_0=\frac{(P^2+q^2-s_0)+\sqrt{(P^2+q^2-s_0)^2+4P^2(m_b^2-q^2)}}{2P^2}.\label{eq:u0}
\end{eqnarray}
Combining Eq. (\ref{sum rules for Ds(2317)}) and Eq. (\ref{QCD
representation of correlation 1}),  we can finally derive the sum
rules for form factors $f_{D_{s0}}^{\ast +}(q^2)$ and
$f_{D_{s0}}^{\ast -}(q^2)$ as
\begin{eqnarray}
 f_{D^*_{s0}}^{+}(q^2)&=&\frac{m_b+m_s}{m^2_{B_s}f_{B_s}}e^{m^2_{B_s}/M^2}\int_{u_0}^1
  \frac{du}{u}[m_b\Phi_{D^*_{s0}}(u)+um_{D^*_{s0}}\Phi_{D^*_{s0}}(u)]e^{-(m_b^2-\bar u q^2+u\bar u
 P^2)/(uM^2)},\nonumber\\
 f_{D^*_{s0}}^{-}(q^2)&=&\frac{m_b+m_s}{m^2_{B_s}f_{B_s}}e^{m^2_{B_s}/M^2}\int_{u_0}^1
  \frac{du}{u}m_{D^*_{s0}}\Phi_{D^*_{s0}}(u)e^{-(m_b^2-\bar u q^2+u\bar u
 P^2)/(uM^2)}.\label{eq:formre1}
\end{eqnarray}

\subsection{Sum rules for $\bar B_s^0\to D_s^+(1968)$ transition form factors}

The form factors responsible for the $\bar B_s^0\to D_s^+$
transition are defined by
\begin{eqnarray}
 \langle D_s^+(P)|\bar c\gamma_{\mu} b|\bar B_s(P+q)\rangle
 &=&f_{D_s}^+(q^2)P_{\mu}+f_{D_s}^-(q^2)q_{\mu}.\label{eq:form2}
\end{eqnarray}
The correlation function associated with the form factors
$f_{D_s}^+(q^2)$ and $f_{D_s}^-(q^2)$ can be chosen as
\begin{eqnarray}
\tilde{ \Pi}_{\mu}(P,q)=-\int d^4x e^{iq\cdot x}\langle
 D^{+}_s(P)|T\{\tilde{j}_{2\mu}(x),j_1(0)\}|0\rangle,\label{eq:cofun2}
\end{eqnarray}
where the current $\tilde{j}_{2\mu}(x)$ is given by
\begin{eqnarray}
\tilde{j}_{2\mu}=\bar c(x)\gamma_{\mu}b(x) .
\end{eqnarray}
One can write the phenomenological representation of the correlation
function at the hadronic level simply by repeating the procedure
given above as
\begin{eqnarray}
\tilde{\Pi} _{\mu }\left( P,q\right)
 &=&\frac{i m^2_{B_s}f_{B_s}}{(m_b+m_s)[m^2_{B_s}-(P+q)^2]}[f_{D_{s}}^+(q^2)P_{\mu}+f_{D_{s}}^-(q^2)q_{\mu}]\nonumber \\
 &&+\int_{s_0^{\bar B_s}}^{\infty}ds
 \frac{\rho_+^h(s,q^2)P_{\mu}+\rho_-^h(s,q^2)q_{\mu}}{s-(P+q)^2}. \label{results for hadronic
level for correlator 2}
\end{eqnarray}
On the other hand, the correlation function at the quark level can
be calculated in the framework of perturbative theory to the leading
order of $\alpha_s$ as
\begin{equation}
\tilde{\Pi}_{\mu}\left(P,q\right) = \int_{u_0}^1 \frac{du}{u}
\frac{i\left[(m_b + um_{D_s})\phi_{D_s}(u)P_{\mu} +
m_{D_s}\phi_{D_s}(u) q_{\mu}\right]}{s-(P+q)^2}, \label{results for
theoretical level for correlator 2}
\end{equation}
where $u_0$ has been  defined in Eq.~(\ref{eq:u0}). Matching the
correlation function obtained in the two different representations
and performing the Borel transformation with respect to the variable
$(P+q)^2$, the sum rules for the form factor $f_{D_s}^+(q^2)$ and
$f_{D_s}^-(q^2)$ can be derived as

\begin{eqnarray}
 f_{D_s}^+(q^2)&=&\frac{m_b+m_s}{m^2_{B_s}f_{B_s}}e^{m^2_{B_s}/M^2}\int_{u_0}^1
  \frac{du}{u}(m_b+um_{D_s})\phi_{D}(u)e^{-(m_b^2-\bar u q^2+u\bar u
 P^2)/(uM^2)},\nonumber\\
 f_{D_s}^-(q^2)&=&\frac{m_b+m_s}{m^2_{B_s}f_{B_s}}e^{m^2_{B_s}/M^2}\int_{u_0}^1
  \frac{du}{u}m_{D_s}\phi_{D}(u)e^{-(m_b^2-\bar u q^2+u\bar u
 P^2)/(uM^2)}. \label{eq:formre2}
\end{eqnarray}

\section{Numerical analysis of sum rules for form factors}

Now we are going to calculate the form factors $f^i_{D_s}(q^2)$ and
$f^i_{D_{s0}}(q^2)$ numerically.  The input parameters used in this
paper \cite{Colangelo:2005hv,
pdg,CKMfitter:2006,Aliev:1983ra,Colangelo:2000dp,ioffe,Azizi:2008tt}
are collected as
\begin{equation}
 \begin{array}{lll}
 |V_{ud}|=0.974,   &|V_{us}|=0.226,    &|V_{ub}|=(3.68^{+0.11}_{-0.08})\times 10^{-3},
\\
|V_{cd}|=0.225,   &|V_{cs}|=0.973,
&|V_{cb}|=(43.99^{+0.69}_{-3.97})\times 10^{-3},
\\
  |V_{td}|=(8.20^{+0.59}_{-0.27})\times 10^{-3}, &|V_{ts}|=40.96\times 10^{-3},   &|V_{tb}|=0.999,\\
 \alpha={(90.6^{+3.8}_{-4.2})}^\circ,& \beta=(21.58^{+0.91}_{-0.81})^{\circ},\;  &
 \gamma=(67.8^{+4.2}_{-3.9})^\circ. \\
 m_b=(4.8\pm 0.1){\rm GeV},\;  &  m_c(1{ \rm GeV})=1.275{\rm GeV},
 & m_s(1\rm{GeV})=142  {\rm MeV},\; \\
m_{B_s}=5.368{\rm GeV},\; &  m_{D_s}=1.968{\rm GeV},\;  &
m_{D^*_{s0}}=2.318{\rm GeV},\;\\
f_{B_s}=(151 \pm 12 ){\rm MeV} \; &  f_{D_s}=(273 \pm 10){\rm
MeV}\,, &  \tilde{f}_{D^*_{s0}}=(225 \pm 25 ){\rm MeV}\,.\;
\end{array}
\end{equation}
As for the decay constant of $B_s$ meson,  we use the results
$f_B=130 {\rm MeV}$ \cite{Aliev:1983ra} and $f_{B_s}/f_{B} = 1.16
\pm 0.09$ \cite{Colangelo:2000dp} determined from QCDSR. The
leptonic decay constants of $D_s(1968)$ and $D_{s0}(2317)$ are
borrowed from Ref. \cite{Colangelo:2005hv,Colangelo:2000dp}. The
threshold parameter $s_0$ can be determined  by demanding the sum
rule results to be relatively stable in allowed region for Borel
mass $M^2$, and its value should be around the mass square of the
first excited states. As for the heavy-light systems, the standard
value of the threshold in the $X$ channel would be
$s^0_{X}=(m_X+\Delta_X)^2$, where $\Delta_X$ is about $0.6$ GeV
\cite{dosch, matheus, bracco, navarra, Wang:2007ys, Colangelo}, and
we simply take it as $(0.6 \pm 0.1)\; \mathrm{GeV}$ corresponding to
$s_0^{B_s}=(36\pm2)\rm{GeV}$ for the error estimation in the
numerical analysis.

With all the parameters listed above,  we can proceed to compute the
numerical values of the form factors. The form factors should not
depend on the the Borel mass $M^2$ in a complete theory.  However,
as we truncate the OPE up to leading conformal spin for the
distribution amplitudes of  $D_s$ meson in the leading Fock
configuration and keep the perturbative expansion in $\alpha_s$ to
the leading order, a manifest dependence of the form factors on the
Borel parameter $M^2$ would emerge. Therefore, one should look for a
working ``window", where the results only vary mildly with respect
to the Borel mass, so that the truncation is acceptable.

\begin{figure}
\begin{center}
\includegraphics[width=7.cm]{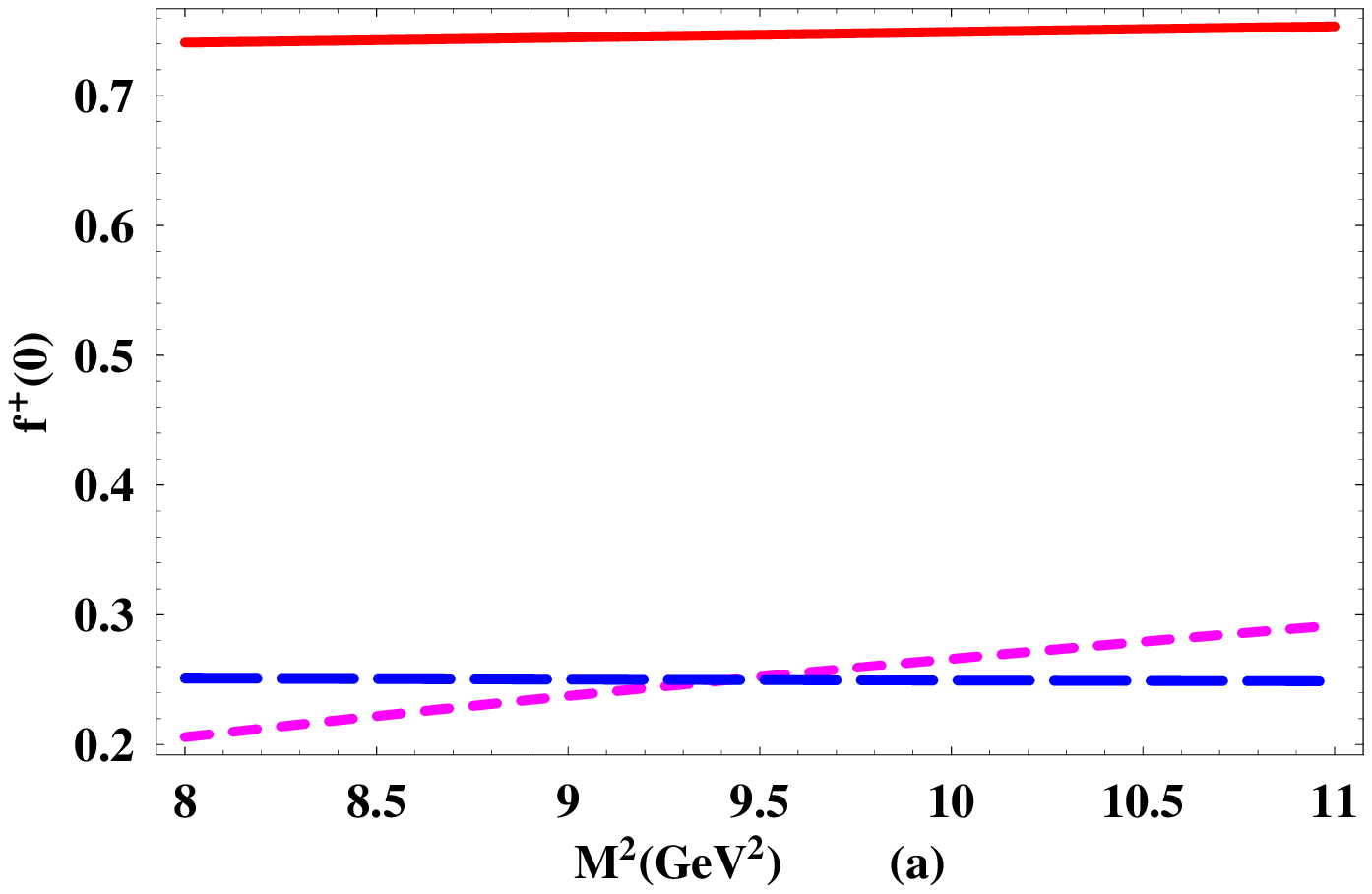}
\hspace{1.0cm}
\includegraphics[width=7.cm]{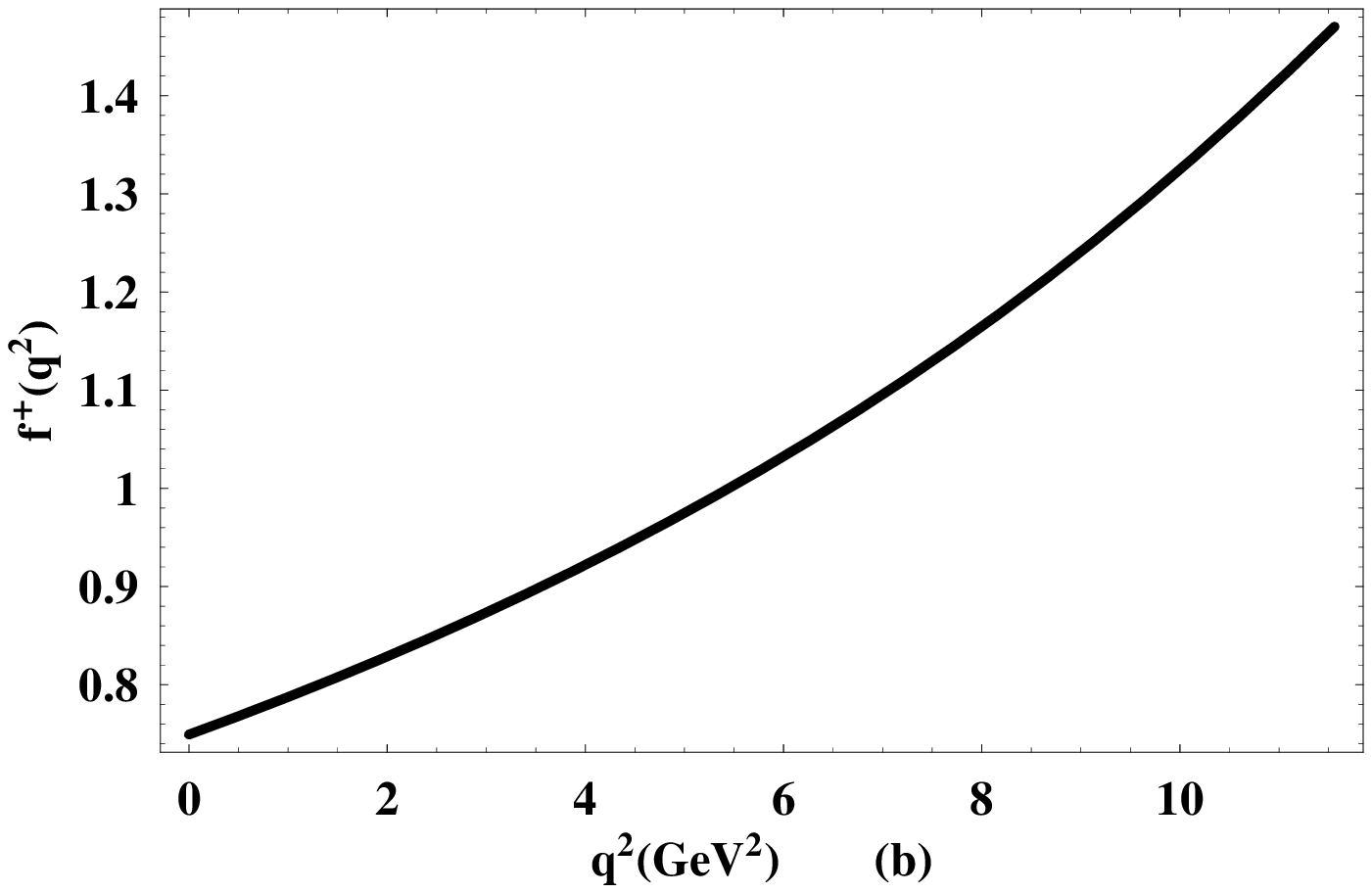}
\vspace{-1.5cm} \caption{Left figure: The dependence form factor
$f_{D_s}^+(0)$ on the Borel mass $M^2$(red solid line), the
contribution of higher states in the whole sum rules(magenta short
dashing line) and the contributions of the twist-3 light cone
distribution amplitude's contribution(blue long dashing line); Right
figure: The dependence of form factor $f_{D_s}^+(q^2)$ on the
momentum transfer $q^2$ within the whole kinematical region.}
 \label{fig:form}
 \end{center}
 \end{figure}

In the first place, we focus on the form factors at zero momentum
transfer. As for the form factor $f_{D_s}^+(0)$ associated with $B_s
\to D_s$ transition,  we require  that the contribution from the
higher resonances and continuum states should be less than 30 \% in
the total sum rules and the value of $f_{D_s}^+(0)$ does not vary
drastically within the selected region for the Borel mass. In view
of these considerations, the Borel parameter $M^2$ should not be too
large in order to insure that the contributions from the higher
states are exponentially damped as can be observed form Eq.
(\ref{eq:formre2}) and the global quark-hadron duality is
satisfactory. On the other hand, the Borel mass could not be too
small for the  validity of OPE near the light-cone for the
correlation function in the deep Euclidean region, since the
contributions of higher twist distribution amplitudes amount to the
higher power of ${1 / M^2}$ to the perturbative part. In this way,
we indeed find a Borel platform $M^2\in [8,11]\rm{GeV}^2$ as plotted
in Fig.~\ref{fig:form}. The value of $f_{D_s}^+(0)$ is
$0.86_{-0.15}^{+0.17}$, where we have combined the uncertainties
from the variation of Borel mass, the fluctuation of threshold
value, the uncertainties of quark masses and the errors of decay
constants for the involved mesons. Following the same procedure, we
can further compute the other form factors numerically, whose
results have been grouped in Table \ref{tab:ffresults}.

Now, we can investigate the $q^2$ dependence of the form factors
$f^i_{D_s}(q^2)$ and $f^i_{D^{\ast}_{s0}}(q^2)$.  It is known that
the OPE for the correlation function (\ref{eq:cofun1},
\ref{eq:form2}) is valid only at small momentum transfer region
$0<q^2<(m_b-m_c)^2-2\Lambda_{\rm{QCD}}(m_b-m_c)$. As for the case
with the large momentum transfer (small recoil region), it is
expected that HQET works well for the $b \to c $ transition. In the
framework of HQET, the matrix elements responsible for  $\bar B_s\to
D_{sx}$ transition can be parameterized as \cite{Kurimoto:2002sb}
\begin{eqnarray}
 \langle D^{*+}_{s0}(P)|\bar c\gamma_{\mu}\gamma_5 b|\bar B_s(P+q)\rangle
 &=&-i\sqrt{m_{B_s}m_{D^*_{s0}}}[\eta_{D^*_{s0}}^+(w)(v+v^{\prime})_{\mu}+\eta_{D^*_{s0}}^-(w)(v-v^{\prime})_{\mu}],\nonumber\\
 \langle D^+_s(P)|\bar c\gamma_{\mu} b|\bar B_s(P+q)\rangle
 &=&\sqrt{m_{B_s}m_{D_{s}}}[\eta_{D_{s}}^+(w)(v+v^{\prime})_{\mu}+\eta_{D_{s}}^-(w)(v-v^{\prime})_{\mu}],\label{eq:formHQET}
\end{eqnarray}
where $v=(P+q)/m_{B_s}$ and $v^{\prime}=P/m_{D_{sx}}$ are the
four-velocity vectors of $B_s$ and $D_{sx}$ mesons with $D_{sx}$
being either $D_s$ or $D^*_{s0}$ meson and $w=v\cdot v^{\prime}$,
Combining Eqs.~(\ref{eq:form1}), (\ref{eq:form2}) and
(\ref{eq:formHQET}), we have
\begin{eqnarray}
 f_i^+(q^2)&=&\frac{1}{\sqrt{m_{B_s}m_{D_{sx}}}}[(m_{B_s}+m_{D_{sx}})\eta_i^+(w)-(m_{B_s}-m_{D_{sx}})\eta_i^-(w)], \nonumber\\
 f_i^-(q^2)&=&\sqrt{\frac{m_{D_{sx}}}{m_{B_s}}}[\eta_i^+(w)+\eta_i^-(w)],\label{eq:feta}
\end{eqnarray}
with $w= (m_{B_s}^2+m_{D_{sx}}^2-q^2 )/ 2m_{B_s}m_{D_{sx}}$. In the
heavy quark limit, the form factors $\eta_{D_{s}}^+(w)$ and
$\eta_{D_{s}}^-(w)$ satisfy the following relations
\begin{eqnarray}
\eta_{D_{s}}^+(w)=\xi(w),  \qquad \eta_{D_{s}}^-(w)=0,
\end{eqnarray}
where $\xi(w)$ is the Isgur-Wise
function\cite{Isgur:1989vq,Isgur:1989ed} with the normalization
$\xi(1)=1$. Similarly, heavy quark symmetry allows to relate the
form factors $\eta_{D^*_{s0}}^{+}(w)$ and $\eta_{D^*_{s0}}^{-}(w)$
to a universal function $\tau_{1/2}(w)$\cite{Isgur:1990jf}
\begin{eqnarray}
\eta_{D^*_{s0}}^{+}(w)+\eta_{D^*_{s0}}^{-}(w)=-2 \tau_{1/2}(w),
\qquad \eta_{D^*_{s0}}^{+}(w)-\eta_{D^*_{s0}}^{-}(w)=2
\tau_{1/2}(w).
\end{eqnarray}
An important relation between the $B \to D^{\ast \ast}$ form factors
at zero recoil region and the slope $\rho^2$ of the $B \to
D^{(\ast)}$ Isgur-Wise function is
\begin{eqnarray}
\rho^2 ={1 \over 4} + \sum_n |\tau_{1/2}^{(n)}|^2 +  \sum_m
|\tau_{3/2}^{(m)}|^2
\end{eqnarray}
under the name of the Bjorken sum rule \cite{Isgur:1990jf}. Here,
$D^{\ast \ast}$ denotes the generic $L=1$ charmed states, the
subscript $n$, $m$ identify the radial excitations of the states
with the same $J^P$. For the $B \to D^{\ast \ast}$  transition form
factors, the essential difference with the Isgur-Wise function
$\xi(y)$ is that one can not invoke heavy quark symmetry arguments
to predict the normalization of $\tau_{1/2}(w)$
\cite{DeFazio:2000up}.

Phenomenologically, one can parameterize the $B_s \to
D_{s}(1968,2317)$ form factors in the small recoil region as
\begin{eqnarray}
\eta_i^{\pm}(w)&=&\eta_i^{\pm}(1)+a_i^{\pm}(w-1)+b_i^{\pm}(w-1)^2,\label{eq:paraHQET}
\end{eqnarray}
where the $\eta_i^{\pm}(w)$ denotes the form factor
$\eta_{D_s}^{\pm}(w)$ and $\eta_{D^*_{s0}}^{\pm}(w)$. The parameters
$\eta_i^{\pm}(1)$, $a_i^{\pm}$ and $b_i^{\pm}$ can be determined
under the condition that the form factors derived in the LCSR and
HQET approaches should be connected in the vicinity of region with
$q^2 \sim (m_b-m_c)^2-2\Lambda_{\rm{QCD}}(m_b-m_c)$. In this way, we
can derive the results of form factors in the whole kinematical
region as shown in Fig. (\ref{fig:form}) as an example.  The values
of all form factors are tabulated in Table \ref{tab:ffresults},
where the results under the QCDSR approaches are also collected for
comparison.

As can be observed from Table \ref{tab:ffresults}, the number of
form factor $\eta_{D^*_{s0}}^+(w)$ at the zero-recoil region
deviates from the zero significantly indicating that the $1/m_c$
power correction is sizeable for the $B \to D^*_{s0}$ transition.
Generally, the expansion of the current
\begin{eqnarray}
\bar{c} \Gamma_i b = \bar{c}_{v_2} \Gamma_i b_{v_1} - {1 \over 2
m_c} \bar{c}_{v_2} \Gamma_i i \! \! \not D_2 b_{v_1} + {1 \over 2
m_b} \bar{c}_{v_2} \Gamma_i i \! \! \not D_1 b_{v_1} + ...
\end{eqnarray}
constitutes  the main source of the power corrections.  
The QCDSR estimation of the form factor  $f^{-}_{D^*_{s0}}$ differs
from that obtained in the LCSR approach in sign implying that the
power corrections  and radiative corrections of correlation function
are in need to reconcile the existing discrepancy between these two
methods.

\begin{table}
\begin{center}
\caption{Numbers of $f_i^{\pm}(0)$ and $\eta_i^{\pm}(w)$ determined
from the LCSR approach, where the uncertainties from the Borel mass,
threshold value, quark masses and decay constants are combined
together. For comparison,  the results estimated in the QCDSR are
also collected  here.} \label{tab:ffresults}
\begin{tabular}{ccc|rcccc}
\hline\hline
\ \ \                     &this work    &QCDSR       $\;\;\;\;\;\;$     &$\eta_i^{\pm}(1)$   &$a_i^{\pm}$       &$b_i^{\pm}$   \\
\hline
\ \ \  $f_{D_{s0}}^+(q^2)$    &$0.53_{-0.11}^{+0.12}$   &$0.40 \pm 0.10$ \cite{Aliev:2006qy}      &$\eta_{D_{s0}}^+(w)$   &$0.45_{-0.10}^{+0.11}$      &$-0.79_{-0.20}^{+0.18}$      &$0.92_{-0.20}^{+0.22}$  \\
\ \ \  $f_{D_{s0}}^-(q^2)$    &$0.18_{-0.04}^{+0.06}$   &$-0.12 \pm 0.13$ \cite{Aliev:2006qy}     &$\eta_{D_{s0}}^-(w)$   &$0.05_{-0.02}^{+0.02}$      &$-0.12_{-0.05}^{+0.04}$      &$0.17_{-0.05}^{+0.06}$  \\
\ \ \  $f_{D_s}^+(q^2)$    &$0.86_{-0.15}^{+0.17}$   &$0.70$\cite{Azizi:2008ty}       &$\eta_{D_s}^+(w)$   &$0.79_{-0.14}^{+0.16}$      &$-1.12_{-0.23}^{+0.20}$      &$0.86_{-0.14}^{+0.16}$  \\
\ \ \  $f_{D_s}^-(q^2)$    &$0.26_{-0.05}^{+0.06}$   &$0.19$\cite{Azizi:2008ty}       &$\eta_{D_s}^-(w)$   &$0.09_{-0.03}^{+0.03}$      &$-0.16_{-0.03}^{+0.04}$      &$0.14_{-0.03}^{+0.04}$  \\
\hline\hline
\end{tabular}
\end{center}
\end{table}

\section{Semileptonic and nonleptonic decays of $\bar  B_s \to D_s(1968,2317)$}

With the form factors derived above, we can further investigate the
semileptonic and nonleptonic decays of $\bar  B_s$ to
$D_s(1968,2317)$ in the factorization approach. Factorization
theorem is a basic theoretical tool to disentangle physical effects
from different energy scales. Factorization in heavy quark decays
was firstly proposed in Ref. \cite{Fakirov:1977ta} as a
phenomenological approximation similar to the vacuum saturation
approximation for the four-quark operator matrix elements.
Intuitively, factorization might work at least to leading-order
approximation, since the partons that eventually  form the emitted
meson escape from the heavy meson remnant as an energetic, low mass,
color singlet system, therefore independently from the remnant.

\subsection{Semileptonic decays of $\bar B_s^0\to D_s^+(1968,2317) l
\bar{\nu}_l$ }

With the free quark amplitude given above and the transition form
factors derived in the LCSR, we  arrive at the differential decay
width for $\bar B_s^0\to D_s^+(1968,2317) l \bar{\nu}_l$ modes
 \begin{eqnarray}
 \frac{d\Gamma}{dq^2}&=&\frac{G_F^2|V_{cb}|^2}{768 \pi^3
 m_B^3}\frac{(q^2-m_l^2)^2}{(q^2)^3}\sqrt{\lambda}
 \bigg[\big(2m_l^2(\lambda+3q^2m_{Dx}^2)+q^2\lambda\big)|f_i^+(q^2)|^2\nonumber\\
 &&+6q^2m_l^2(m_B^2-m_{Dx}^2-q^2)f_i^+(q^2)f_i^-(q^2)+6q^4m_l^2|f_i^-(q^2)|^2\bigg],\label{eq:semil}
 \end{eqnarray}
with $\lambda=(m_B^2-m_{Dx}^2-q^2)^2-4q^2m_{Dx}^2$.

 \begin{figure}
 \begin{center}
 \includegraphics[width=7.cm]{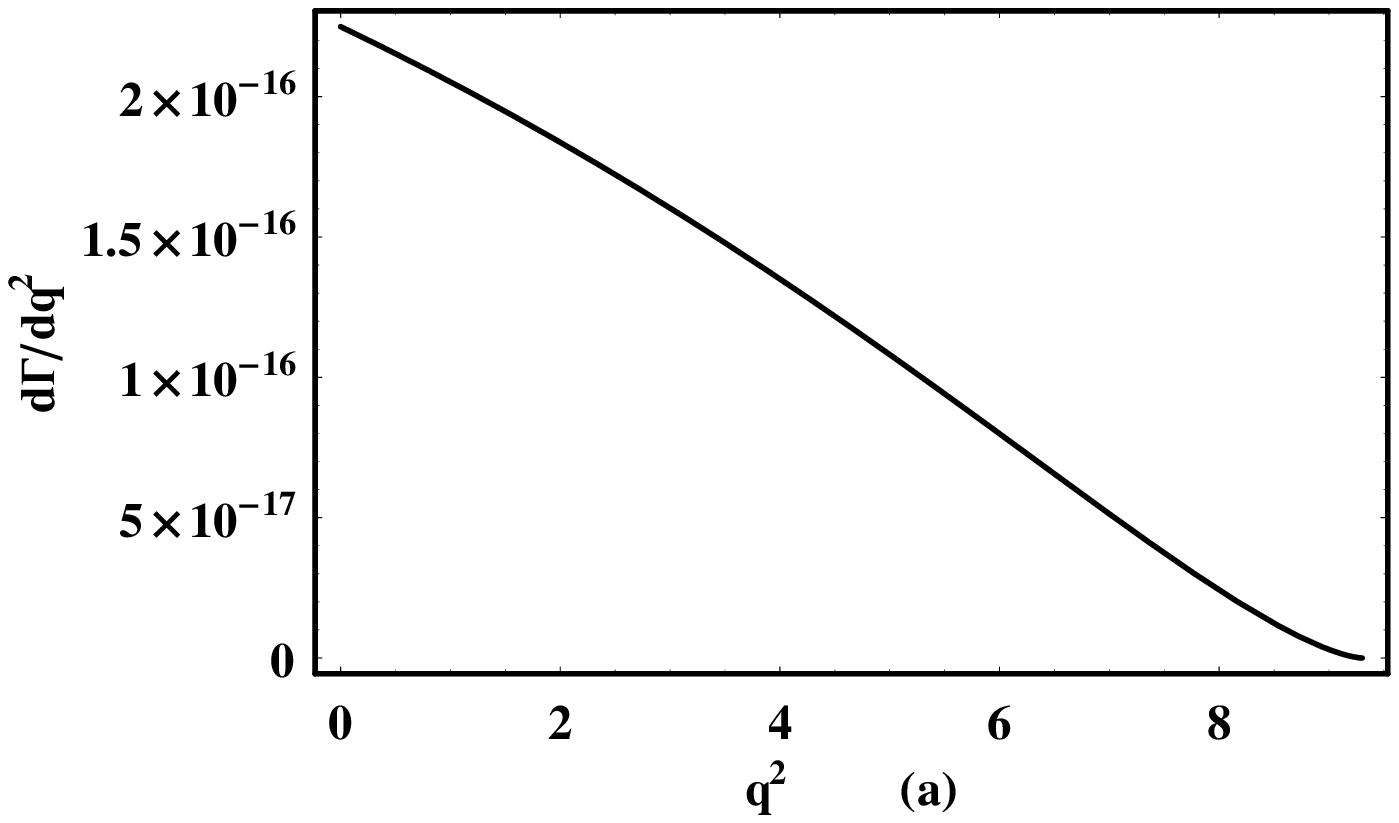}
 \hspace{1.0cm}
 \includegraphics[width=7.cm]{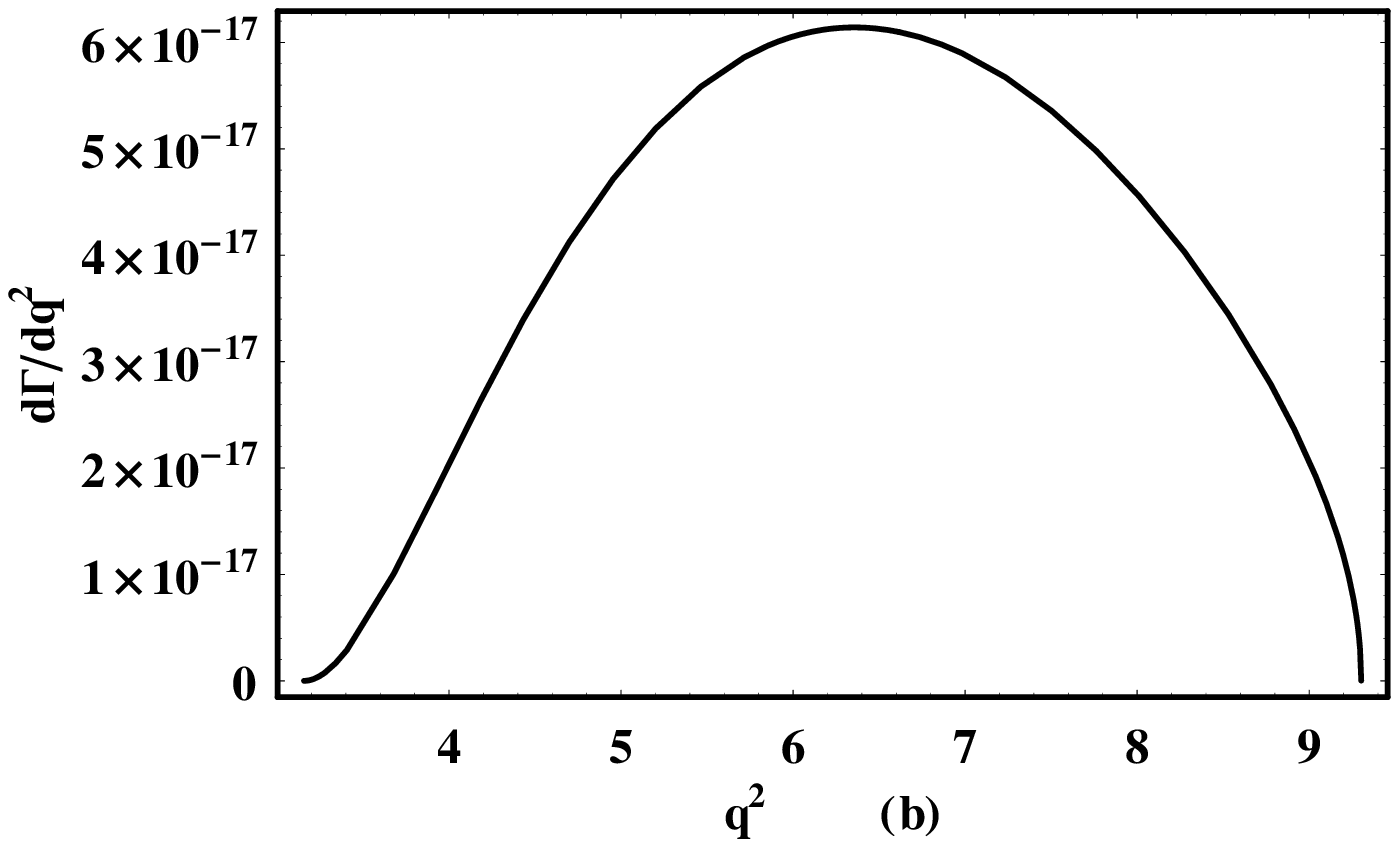}
 \vspace{-1.5cm}
\caption{The $q^2$ dependence of differential decay width
$\frac{d}{dq^2} \Gamma(\bar B_s^0\to D_{s0}^+l^-\bar\nu_l)$ for the
final states with  $l=e,\mu$ (left figure)  and $l=\tau$ (right
figure).}
 \label{fig:semiDs1}
 \end{center}
 \end{figure}

 \begin{figure}
 \begin{center}
 \includegraphics[width=7.cm]{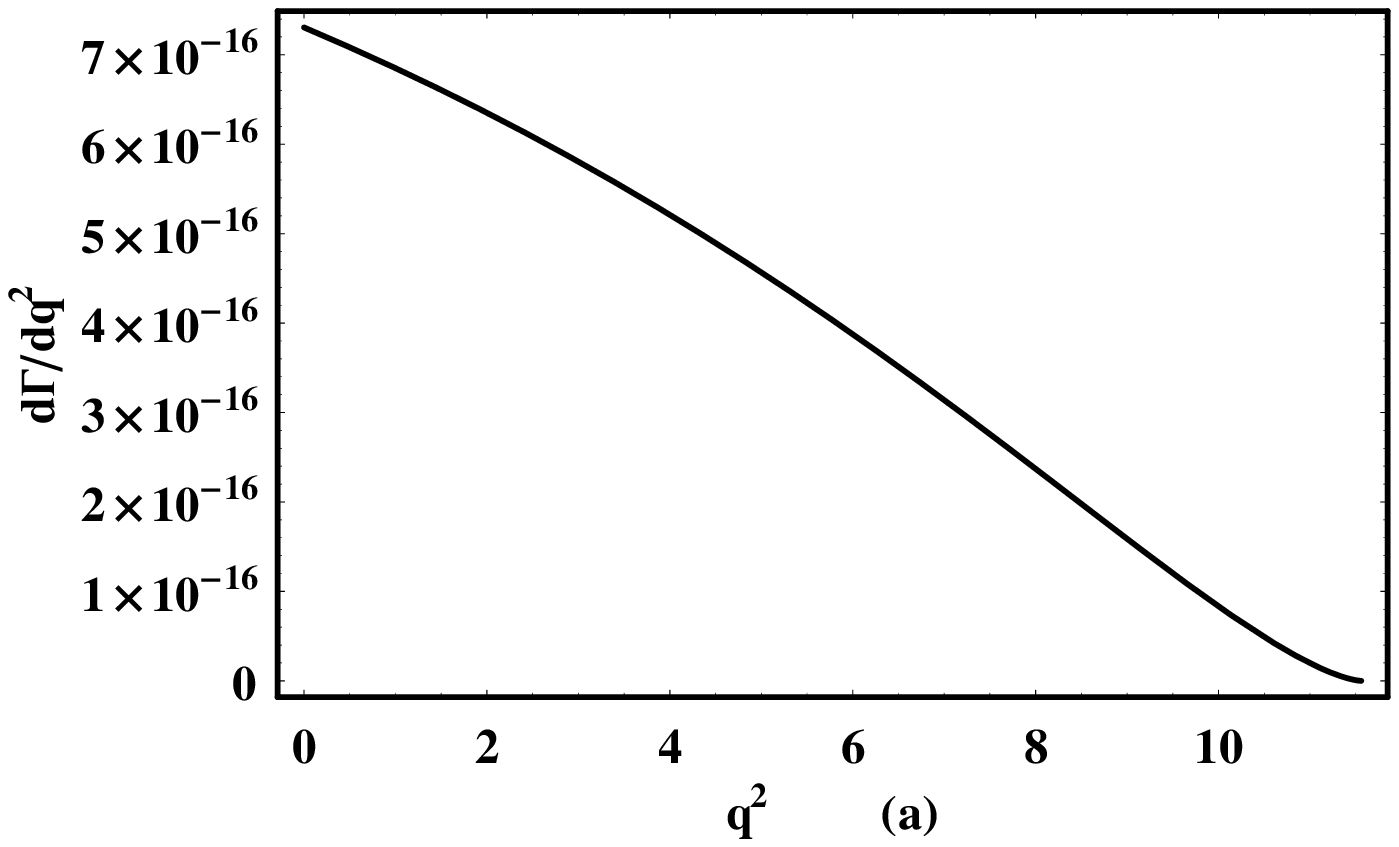}
 \hspace{1.0cm}
 \includegraphics[width=7.cm]{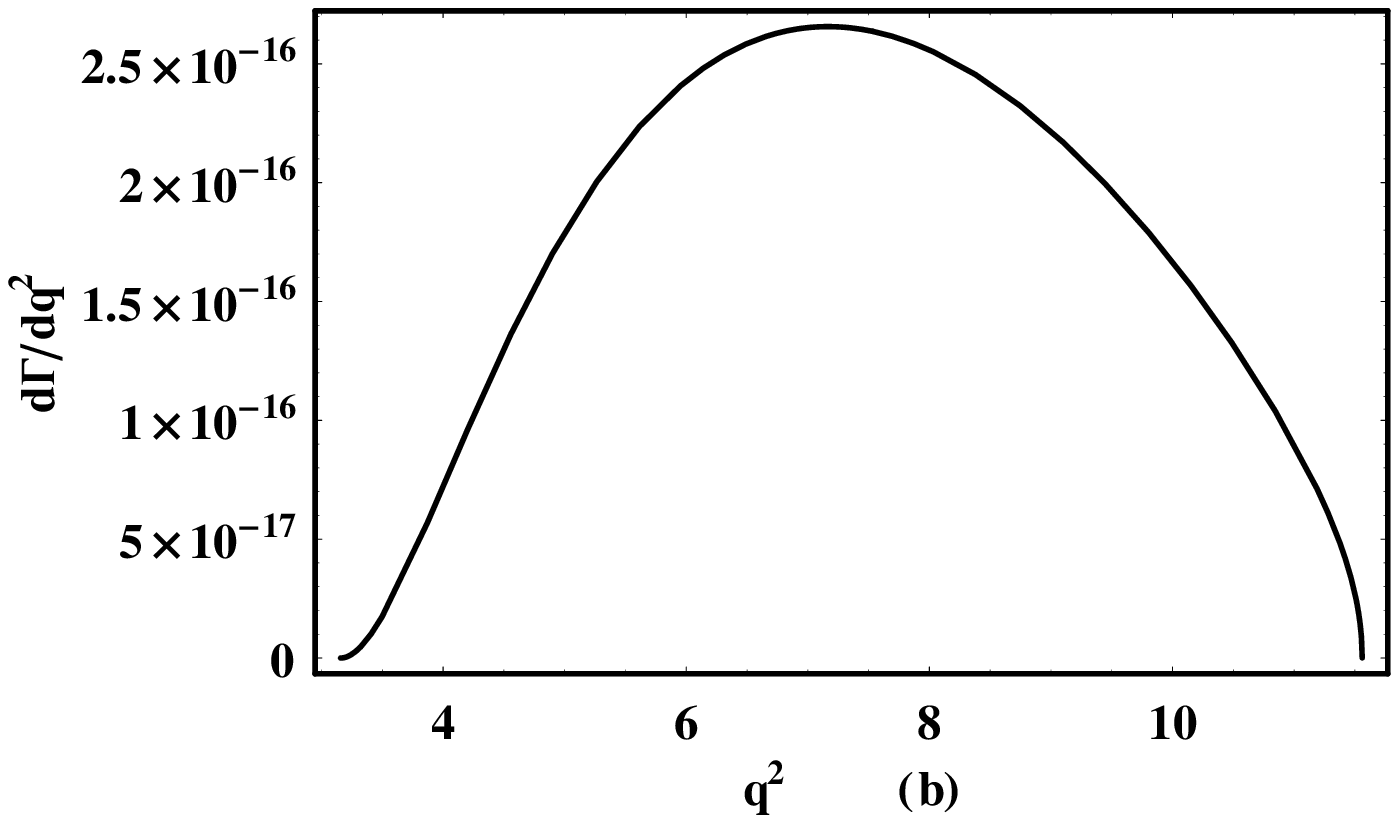}
 \vspace{-1.5cm}
\caption{The $q^2$ dependence of differential decay width
$\frac{d}{dq^2} \Gamma(\bar B_s^0\to D_{s}^+l^-\bar\nu_l)$ for the
final states with  $l=e,\mu$ (left figure)  and $l=\tau$ (right
figure).}
 \label{fig:semiDs2}
 \end{center}
 \end{figure}

For convenience, the $q^2$ dependence of these invariant functions
are also plotted in Fig. \ref{fig:semiDs1} and \ref{fig:semiDs2}.
Integrating Eq. (\ref{eq:semil}), we get the branching fractions of
$\bar B_s^0\to D_s^+(1968,2317) l \bar{\nu}_l$ as grouped in Table
\ref{tab:Brforsemi}. It can be observed from this table that the
orders of magnitudes for $BR(\bar B_s^0\to D_s^+(1968,2317) l
\bar{\nu}_l)$ obtained in the quark model and sum rule approaches
are consistent with each other. Besides, we can also find that the
decay rates for the  final state with $\tau$ lepton are generally
$3-4$ times smaller than those for the muon case due to the
suppression of phase spaces. Once the  data on the $\bar B_s^0\to
D_s^+(2317) l \bar{\nu}_l$ are available, the theoretical
predictions presented here can be put to the experimental scrutiny
to test the ordinary $c \bar{s}$ picture of $D_s(2317)$ meson.

 \begin{table}
 \caption{Branching ratios for the semileptonic decays  $\bar B_s^0\to D_s^+(1968,2317) l
\bar{\nu}_l$  with the form factors estimated in LCSR, where the
results calculated in constituent quark model and QCDSR are also
displayed for comparison.}
 \label{tab:Brforsemi}
 \begin{center}
 \begin{tabular}{c c c}
 \hline\hline
 \ \ \ $\bar B_s^0\to D_{s0}^+l^-\bar\nu_l$         &$l=e,\mu$                             &$l=\tau$ \\
 \ \ \  this work                           &$(2.3_{-1.0}^{+1.2})\times10^{-3}$            &$(5.7_{-2.3}^{+2.8})\times10^{-4}$\\
 \ \ \  QCDSR\cite{Aliev:2006qy}    &$\sim 10^{-3}$                                &$\sim 10^{-4}$  \\
 \ \ \  Constituent Quark Model\cite{Zhao:2006at}   &$(4.90-5.71)\times 10^{-3}$           &$$  \\
 \ \ \ QCDSR in HQET\cite{Huang:2004et}      &$(0.9-2.0)\times 10^{-3} $                                 &$$\\
 \hline\hline
 \ \ \  $\bar B_s^0\to D_s^+l^-\bar\nu_l$        &$l=e,\mu$            &$l=\tau$ \\
 \ \ \  this work       &$(1.0_{-0.3}^{+0.4})\times10^{-2}$            &$(3.3_{-1.1}^{+1.4})\times10^{-3}$\\
 \ \ \  Constituent Quark Model\cite{Zhao:2006at}   &$(2.73-3.00)\times 10^{-2}$           &$$  \\
 \ \ \  QCDSR \cite{Azizi:2008tt}            &$(2.8-3.8)\times 10^{-2}$            &$$  \\
 \hline\hline
 \end{tabular}
 \end{center}
 \end{table}

 \subsection{Nonleptonic decays of $\bar{B}_s \to  D_s^+(1968,2317) M$ }

Now, we turn to the calculations of nonleptonic decays $\bar{B}_s
\to D_s^+(1968,2317) M$, where $M$ can be a light meson or a charmed
meson. As mentioned above, the factorization assumption will be
employed to decompose the matrix element of four-quark operator
\begin{eqnarray}
\langle D_{sx} M|Q|\bar B_s\rangle=\langle M|j_2|0\rangle \langle
 D_{sx} |j_1|\bar B_s\rangle, \label{eq:NF}
\end{eqnarray}
into the the $\bar{B}_s \to  D_s^+(1968,2317)$  transition form
factors and the decay constant of $M$.

For a light meson $M$, only the tree-operators in Eq. (\ref{HeffSM})
can contribute to these decay modes induced by the $b \to c$
transition. Then, the decay width for $\bar{B}_s \to
D_s^+(1968,2317) L$ can be written as
\begin{eqnarray}
 \Gamma(\bar B_s\to D_{sx}^+L^-)&=&\frac{G_F^2|\vec{P}|}{16\pi m^2_{B_s}}
 |V_{cb}V^*_{uq}a_2(\mu)|^2 f_L^2\times \bigg \{
  \begin{array}{cc}
     |f_{D_{sx}}^+(m^2_L) m_L (\epsilon^* \cdot P)  |^2 \hspace{3.0 cm} (\mbox{$L$=$V$}), \\
     |\frac{m^2_{B_s}-m^2_{D_{sx}}-m^2_L}{2}f_{D_{sx}}^+(m^2_L)+f_{D_{sx}}^-(m^2_L)m^2_L |^2\;\;
    \hspace{0.2 cm} (\mbox{$L$=$S, \,\, P$}),
  \end{array}\label{eq:GammaforL}
\end{eqnarray}
where $L$ denotes a light meson; $V$, $P$ and $S$ label the vector,
pseudoscalar and scalar mesons respectively. The magnitude of the
three-momentum for the recoiled charmed meson  is
\begin{eqnarray}
|\vec{P}|=\frac{[(m^2_{B_s}-(m_{D_{sx}}+m_L)^2)(m^2_{B_s}-(m_{D_{sx}}-m_L)^2)]^{1/2}}{2m_{B_s}},\label{eq:vecP}
\end{eqnarray}
and the  decay constant $f_L$ has been collected in
Table~\ref{tab:decayconstants}. The   decay constants for the  light
pseudoscalar mesons  are taken from the Particle Data
Group~\cite{pdg} and the vector meson longitudinal decay constants
are extracted from the data on $\tau^- \to (\rho^-,K^{*-})
\nu_\tau$. To determine the decay constants for the scalar meson
$D_0^{\ast}$ and vector meson $D_{s}^{\ast}$, the following relation
\begin{eqnarray}
  {f_{D_{s0}^{\ast}} \over f_{D_0^{\ast}}} \approx {
f_{D_{s}^{\ast}} \over f_{D^{\ast}}} \approx { f_{B_s} \over f_{B}}
\end{eqnarray}
is assumed in this work. The decay constant of vector meson
$D^{\ast}$ is borrowed from Ref. \cite{Colangelo:2000dp}. The energy
scale of the Wilson coefficient $a_2(\mu)$ is varied from $0.5 m_b$
to $1.5 m_b$ in the error estimations.

 \begin{table}
 \caption{Decay constants of light and charmed  mesons (unit: $\rm{MeV}$).}
 \label{tab:decayconstants}
 \begin{center}
 \begin{tabular}{c c c c c c c c c c}
 \hline\hline
$f_{\pi}$   & \hspace {1 cm}  $f_{K}$   & \hspace {1 cm}  $f_{\rho}$  &\hspace {1 cm}  $f_{K^*}$      & \hspace {1 cm}  $f_{D}$   &\hspace {1 cm}  $f_{D_0^*}$   &\hspace {1 cm}  $f_{D^*}$ &\hspace {1 cm}  $f_{D_s^*}$ \\
 $131$        & \hspace {1 cm}  $160$    & \hspace {1 cm}  $209 \pm 2$        & \hspace {1 cm} $217 \pm 5$         & \hspace {1 cm}  $206 \pm 9$     & \hspace {1 cm} $95 \pm 10$         & \hspace {1 cm}  $270 \pm 35$    & \hspace {1 cm}  $312 \pm 40$     \\
 \hline\hline
 \end{tabular}
 \end{center}
 \end{table}

Substituting the form factors obtained in the previous sections into
Eq.~(\ref{eq:GammaforL}),  we can get the decay rates of $\bar{B}_s
\to D_s^+(1968,2317) L$ as shown in Table \ref{tab:Brfornon1}. From
this table, the results evaluated in the factorization approach are
in accord with that predicted in the PQCD approach and the available
data, which implies that the factorization assumption (FA) works
well for these color allowed modes as expected.
\begin{table}
 \caption{Branching ratios(unit: $10^{-4}$) for the nonleptonic decays
 $\bar B_s\to D_{sx}^+L$ ($L$ denotes  a light meson) estimated in the FA
 with the form factors obtained in the LCSR, where we have combined the
 uncertainties from the form factors, the scale-dependence and CKM matrix elements.
 In Ref. \cite{Azizi:2008ty}, the authors also
 employ the naive factorization but take the transition form factors
calculated in the three-point  sum rules.}
 \label{tab:Brfornon1}
 \begin{center}
 \begin{tabular}{ccccc}
 \hline\hline
 Channels        & this work       &PQCD\cite{Li:2008ts}     & Exp.\cite{pdg}      & FA \cite{Azizi:2008ty}      \\
 \hline
 $\bar B_s^0 \to D^{\ast +}_{s0}\pi^-$        &$5.2_{-2.1}^{+2.5}$      &$$        &$$    &$$    \\
 $\bar B_s^0 \to D^{\ast +}_{s0}K^-$          &$0.4_{-0.2}^{+0.2}$      &$$        &$$    &$$     \\
 $\bar B_s^0 \to D^{\ast +}_{s0}\rho^-$       &$13_{-5}^{+6}$           &$$        &$$    &$$     \\
 $\bar B_s^0 \to D^{\ast +}_{s0}K^{*-}$       &$0.8_{-0.3}^{+0.4}$      &$$        &$$    &$$    \\
 \hline
 $\bar B_s^0 \to D^+_s\pi^-$        &$17_{-6}^{+7}$         &$19.6_{-7.5-6.2-0.6}^{+10.6+6.3+0.6}$     &$32\pm9$    &$11.2\pm 22.7$    \\
 $\bar B_s^0 \to D^+_sK^-$          &$1.3_{-0.4}^{+0.5}$      &$1.70_{-0.66-0.56-0.05}^{+0.87+0.53+0.05}$    &$$    &$0.83\pm 2.51$  \\
 $\bar B_s^0 \to D^+_s\rho^-$       &$42_{-14}^{+17}$         &$47.0_{-17.7-14.8-1.37}^{+24.9+15.3+1.36}$     &$$    &$$    \\
 $\bar B_s^0 \to D^+_sK^{*-}$       &$2.4_{-0.8}^{+1.0}$      &$2.81_{-1.09-0.85-0.09}^{+1.47+0.79+0.09}$     &$$    &$0.60\pm 2.82$ \\
 \hline\hline
 \end{tabular}
 \end{center}
 \end{table}

Moreover, it is also helpful to define the following ratios
\begin{eqnarray}
R_1 & \equiv  & { BR(\bar B_s^0 \to D^{\ast +}_{s0}\pi^-) \over
BR(\bar B_s^0 \to D^{\ast +}_{s0}K^-) } \approx { BR(\bar B_s^0 \to
D^+_s\pi^-) \over BR(\bar B_s^0 \to D^+_s K^-)} \approx \bigg | {
V_{ud} \over  V_{us}} { f_{\pi} \over f_K }\bigg |^2 \approx 12.6 ,
\nonumber  \\
R_2 & \equiv & { BR(\bar B_s^0 \to D^{\ast +}_{s0}\rho^-) \over
BR(\bar B_s^0 \to D^{\ast +}_{s0}K^{*-})} \approx { BR(\bar B_s^0
\to D^+_s\rho^-) \over  BR(\bar B_s^0 \to D^+_sK^{*-})} \approx
\bigg | { V_{ud} \over  V_{us}} { f_{\rho} \over f_{K^{\ast}} }\bigg
|^2 \approx 17.4 ,
\end{eqnarray}
which are consistent with those collected in Table
\ref{tab:Brfornon1}.

As for the two charmed meson decays of $B_s$ meson, the decay width
in the factorization approach can be given by
\begin{eqnarray}
 \Gamma(\bar B_s\to D_{sx}^+X)&=&\frac{G_F^2|\vec{P}|}{16\pi m^2_{B_s}}
 |V_{cb}V^*_{cq}a_2(\mu)-V_{tb}V^*_{tq}[a_4(\mu)+a_{10}(\mu)+r_q (a_6(\mu)+a_8(\mu))]|^2 f_X^2\nonumber\\
&&\times \bigg \{
  \begin{array}{cc}
     |f_{D_{sx}}^+(m^2_X) m_X (\epsilon^* \cdot P)  |^2 \hspace{3.8 cm} (\mbox{$X$=$V$}), \\
     |\frac{m^2_{B_s}-m^2_{D_{sx}}-m^2_X}{2}f_{D_{sx}}^+(m^2_X)+f_{D_{sx}}^-(m^2_X)m^2_X |^2\;\;
    \hspace{1 cm} (\mbox{$X$=$S, \,\, P$}),
  \end{array}\label{eq:GammaForD}
\end{eqnarray}
with
\begin{eqnarray}
 r_q=\left\{
 \begin{array}{rrrrr}
 {2m_X^2 \over (m_b-m_c)(m_c+m_q)}   \hspace{1.0cm} (\bar B_s^0\to D_s^+X,\;\mbox{$X$=$P$}),\\
 {2m_X^2 \over (m_b-m_c)(m_c-m_q)}   \hspace{1.0cm} (\bar B_s^0\to D_s^+X,\;\mbox{$X$=$S$}),\\
 -{2m_X^2 \over (m_b-m_c)(m_c-m_q)}   \hspace{1.0cm} (\bar B_s^0\to D_{s0}^{*+}X,\;\mbox{$X$=$P$}),\\
 -{2m_X^2 \over (m_b-m_c)(m_c-m_q)}   \hspace{1.0cm} (\bar B_s^0\to D_{s0}^{*+}X,\;\mbox{$X$=$S$}),\\
 0   \hspace{2.5cm} (\bar B_s^0\to D_{sx}^{+}X,\;\mbox{$X$=$V$}),\\
 \end{array}
 \right \}
\end{eqnarray}
where the quark masses in the above equation are the current quark
masses.

Combining the  Eq. (\ref{eq:GammaForD}) and the form factors listed
above, one can easily get the branching ratios of $\bar B_s\to
D_{sx}^+X$ ($X$ being a charmed meson) as shown in Table
\ref{tab:Brfornon2}. As can be seen from this table, the decay model
$\bar B_s^0 \to D^+_{s0}D_s^-$ possesses  a quite large branching
ratio of order $10^{-2}$, which should be detected  easily at the
large colliders such as  Tevatron  and LHC. Moreover, the
theoretical predictions on the  $\bar B_s^0 \to D^+_s\pi^-$ decay
can accommodate the experimental data within the error bars. As for
the $\bar B_s^0 \to D^+_s D_s^{*-}$ decay,  only the upper bound for
this mode is available at present, which is  also  respected by our
predictions.

Subsequently,  the ratio of decay rates between the Cabibbo favored
and suppressed modes can be estimated as
\begin{eqnarray}
R_3 & \equiv & {BR(\bar B_s^0 \to D^{\ast +}_{s0}D_s^-) \over
BR(\bar B_s^0 \to D^{\ast +}_{s0}D^-)} \approx   {BR(\bar B_s^0 \to
D^+_s D_s^-) \over  BR(\bar B_s^0 \to D^+_s D^-)} \approx \bigg |{
V_{cs} \over V_{cd}  } { f_{D_s^-}\over
f_{D^-}} \bigg |^2 , \nonumber \\
R_4 & \equiv &  {BR(\bar B_s^0 \to D^{\ast +}_{s0}D_s^{\ast -})
\over BR(\bar B_s^0 \to D^{\ast +}_{s0}D^{\ast -})} \approx {BR(\bar
B_s^0 \to D^+_s D_s^{\ast -}) \over BR(\bar B_s^0 \to D^+_s D^{\ast
-})} \approx \bigg |{ V_{cs} \over V_{cd}  } {
f_{D_s^{\ast -}}\over f_{D^{\ast -}}} \bigg |^2 , \nonumber \\
R_5 & \equiv &  { BR(\bar B_s^0 \to D^{\ast +}_{s0}D_{s0}^{\ast -})
\over BR(\bar B_s^0 \to D^{\ast +}_{s0}D_0^{*-})} \approx { BR(\bar
B_s^0 \to D^+_s D_{s0}^{\ast -}) \over BR(\bar B_s^0 \to D^+_s
D_0^{*-}) } \approx \bigg |{ V_{cs} \over V_{cd}  } {
f_{D_{s0}^{\ast -}}\over f_{D_{0}^{\ast -}}} \bigg |^2 ,
\end{eqnarray}
in the naive factorization  without the  contributions from penguin
operators.  Such naive estimations are in good agreement with that
presented in Table {\ref{tab:Brfornon2}}, which also indicates that
the two charmed-meson decays of $B_s$ meson  governed  by the $b \to
q c \bar{c}$ ($q=s\,\, d$) transition are dominated by the tree
operators.

 \begin{table}
 \caption{Branching ratios(unit: $10^{-3}$) for the nonleptonic decays
  $\bar B_s\to D_{sx}^+X$ ($X$ denotes a charmed meson) estimated in the FA
 with the form factors obtained in the LCSR, where the
 uncertainties from the form factors, the scale-dependence
 and CKM matrix elements have been combined together.}
 \label{tab:Brfornon2}
 \begin{center}
 \begin{tabular}{ccc}
 \hline\hline
 Channels         & This work        &Exp. \cite{pdg}       \\
 \hline
 $\bar B_s^0 \to D^{\ast +}_{s0}D_s^-$          &$13_{-5}^{+7}$       &$$    \\
 $\bar B_s^0 \to D^{\ast +}_{s0}D^-$          &$0.5_{-0.2}^{+0.2}$    &$$    \\
 $\bar B_s^0 \to D^{\ast +}_{s0}D_{s0}^-$       &$2.1_{-0.8}^{+1.0}$  &$$    \\
 $\bar B_s^0 \to D^{\ast +}_{s0}D_0^{*-}$     &$0.2_{-0.1}^{+0.1}$    &$$    \\
 $\bar B_s^0 \to D^{\ast +}_{s0}D_s^{*-}$       &$6.0_{-2.4}^{+2.9}$  &$$    \\
 $\bar B_s^0 \to D^{\ast +}_{s0}D^{*-}$       &$0.2_{-0.1}^{+0.1}$    &$$    \\
 \hline
 $\bar B_s^0 \to D^+_s D_s^-$            &$35_{-12}^{+14}$     &$110\pm40$  \\
 $\bar B_s^0 \to D^+_s D^-$            &$1.1_{-0.4}^{+0.4}$    &$$    \\
 $\bar B_s^0 \to D^+_s D_{s0}^{\ast -}$         &$5.3_{-1.8}^{+2.2}$  &$$   \\
 $\bar B_s^0 \to D^+_s D_0^{*-}$       &$0.2_{-0.1}^{+0.1}$    &$$   \\
 $\bar B_s^0 \to D^+_s D_s^{*-}$         &$33_{-11}^{+13}$     &$<121$    \\
 $\bar B_s^0 \to D^+_s D^{*-}$         &$1.4_{-0.5}^{+0.6}$    &$$    \\
 \hline\hline
 \end{tabular}
 \end{center}
 \end{table}

\section{Discussion and conclusion}

A detailed analysis of properties about the new charming mesons such
as $D_s(2317)$ has become a prominent part of the ongoing and
forthcoming experimental programs at various facilities worldwide.
The production characters of charmed mesons in the $B_s$ decays are
especially interesting for highlighting the understanding of QCD
dynamics and enriching our knowledge of flavor physics. More
importantly, the theory underlying the description of the decays
induced by the $b \to c$ transition is mature currently. In view of
the large mass of $b$ quark, heavy quark expansion works well enough
to enable a precise determination of the decay amplitude.

LCSR approach is employed to compute the $\bar B_s^0\to
D_s^+(1968,2317)$ transition form factors at the large recoil region
and the results are then extended to the  small recoil region in the
framework of HQET.  Our results show that the power correction to
the form factor $\eta_{D_{s}}^+(w)$ responsible for the $\bar
B_s^0\to D_s^+(1968)$ at the zero-recoil region transition is
numerically small, since this form factor only receive the
corrections at order $1/m_{b}^2$ and $1/m_{c}^2$ as indicated by the
Luke's theorem \cite{Luke}. However, the power correction to the
form factor $\eta_{D_{s0}^{\ast}}^{ +}(w)$ relevant to the $B \to
D_{s0}^{\ast +}$ transition is sizeable.

Subsequently, we utilize the form factors estimated in the LCSR
approach to perform a careful study on  the semileptonic decays
$\bar B_s^0\to D_s^+(1968,2317) l \bar{\nu}_l$.  It has been shown
in this work that the branching fraction of the semileptonic $\bar
B_s^0\to D_s^+(2317) \mu \bar{\nu}_\mu$ decay is around $2.3 \times
10^{-3}$, which should be detectable  with ease at the Tevatron and
LHC. The decay rates of semileptonic modes for the final states with
$\tau$ lepton are approximately $3 - 4$ times smaller than those
with muon due to the suppression of phase spaces. In addition, the
branching fractions of $\bar B_s^0\to D_s^+(1968) l \bar{\nu}_l$ are
almost one order large than that for the $B_s^0\to D_s^+(2317) l
\bar{\nu}_l$ decays.

Nonleptonic decays $B_s \to D_s^+(1968,2317)  M$  are also
investigated in the framework of factorization approach in this
work. It is found that the theoretical predictions  for $B_s \to
D_s^+(1968,2317) L$ presented here are in agreement with those
obtained in the $k_T$ factorization, which supports the success  of
color transparence mechanism in the color allowed decay modes.
Moreover, $\bar B_s^0 \to D^{\ast +}_{s0}D_s^-$ owns a quite large
branching ratio as $1.3 \times 10^{-2}$, which should be accessible
experimentally. More theoretical results worked out here are
expected to be tested by the large colliders in the near future.

\section*{Acknowledgement}
This work is partly supported by National Natural Science Foundation
of China under the Grant No. 10735080, 10625525, and 10525523. The
authors would like to thank K. Azizi, M.Q. Huang and H.n. Li for
helpful discussions.


\end{document}